\documentclass[sigconf, authorversion, nonacm]{acmart}

\AtBeginDocument{%
  }

\acmConference[Workshop at CHI 2024]{Workshop on Theory of Mind in Human-AI Interaction at CHI 2024}{May 12, 2024}{Honolulu, Hawaii, USA}

\begin{document}

\title[Inferring Human Intentions]{Inferring Human Intentions from Predicted Action Probabilities}

\author{Lei Shi}
\affiliation{
\institution{University of Stuttgart}
\country{Germany}
}
\email{lei.shi@vis.uni-stuttgart.de}

\author{Paul-Christian Bürkner}
\affiliation{
\institution{University of Stuttgart}
\country{Germany}
}
\email{paul.buerkner@gmail.com}

\author{Andreas Bulling}
\affiliation{
\institution{University of Stuttgart}
\country{Germany}
}
\email{andreas.bulling@vis.uni-stuttgart.de}

\renewcommand{\shortauthors}{Shi et al.}

\begin{abstract}
  Inferring human intentions is a core challenge in human-AI collaboration but while Bayesian methods struggle with complex visual input, deep neural network (DNN) based methods do not provide uncertainty quantifications.
In this work we combine both approaches for the first time and show that the predicted next action probabilities contain information that can be used to infer the underlying user intention.
We propose a two-step approach to human intention prediction:
While a DNN predicts the probabilities of the next action, MCMC-based Bayesian inference is used to infer the underlying intention from these predictions.
This approach not only allows for the independent design of the DNN architecture
but also the subsequently fast, design-independent inference of human intentions.
We evaluate our method using a series of experiments on the Watch-And-Help (WAH) and a keyboard and mouse interaction dataset. 
Our results show that our approach can accurately predict human intentions from observed actions
and the implicit information contained in next action probabilities.
Furthermore, we show that our approach can predict the correct intention even if only a few actions have been observed.

\end{abstract}






\received{22 February 2024}
\received[accepted]{21 March 2024}

\maketitle

\section{Introduction}
\label{sec:intro}

\begin{figure}[t]
    \centering
    \includegraphics[width=\linewidth]{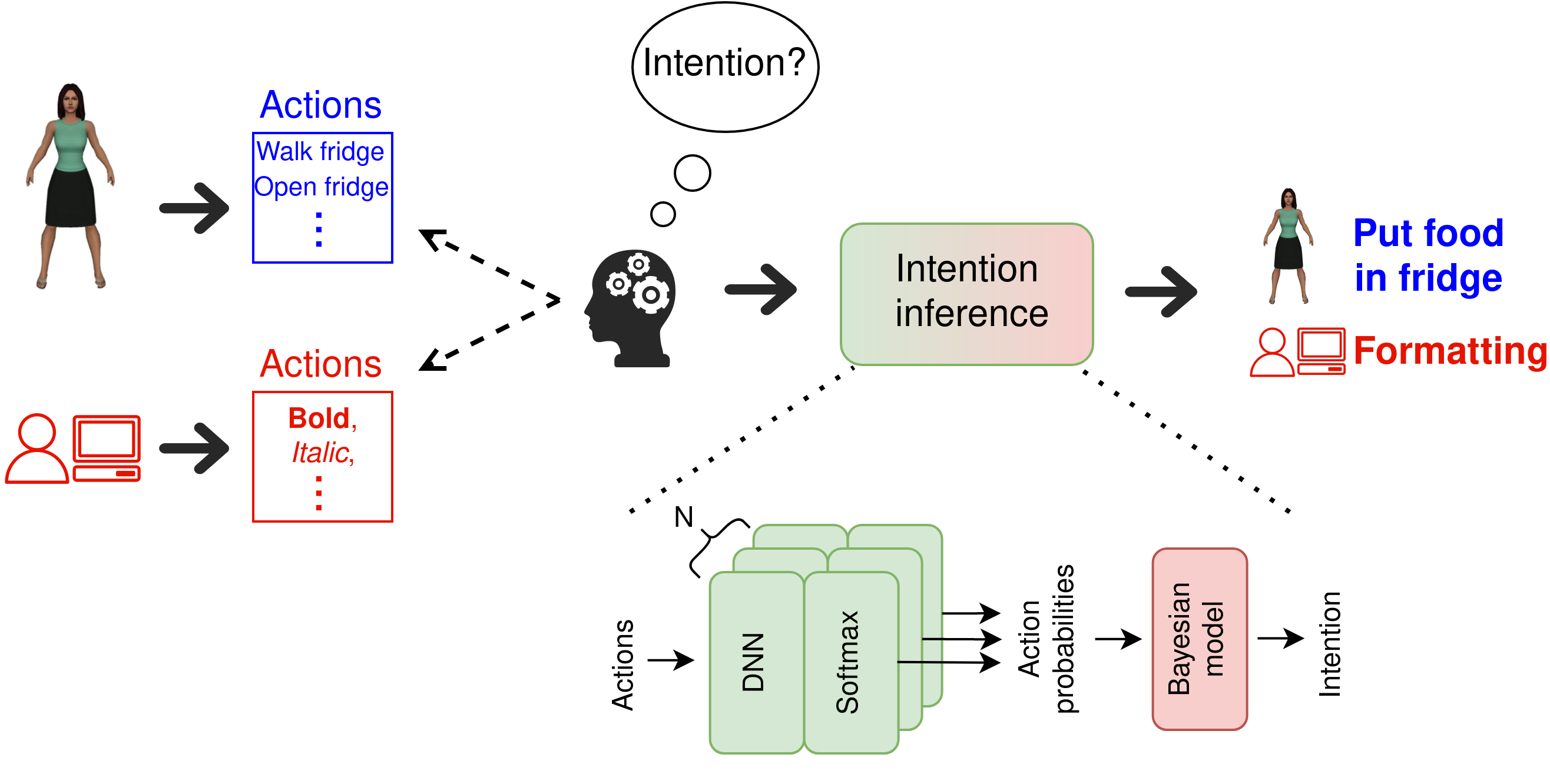}
    \caption{Overview of our proposed method to predict human intentions. An agent observes the human actions and tries to infer the human's intention. Deep Neural Networks (DNNs) together with a Bayesian model infers the human intention.    }
    \label{fig:teaser}
\end{figure}

A hallmark of human cognition is Theory of Mind (ToM), i.e. our ability to attribute mental states to others, such as thoughts, beliefs, or feelings. A critical requirement is the ability to understand others’ intentions, i.e. their commitment to carrying out a particular action in the future \cite{malle2001intentions}. This understanding enables us to anticipate others’ actions \cite{huang2015using} and is thus essential for us to engage in social communication and to interact naturally, effortlessly, and seamlessly with each other.
In contrast, despite its importance for the research on human-computer interaction (HCI) and human-AI collaboration, 
current AI agents still lack the ability of ToM and fail to understand users’ attention, predict their intentions, and anticipate their needs and actions. This limits the agents to operating after users' actions, thereby drastically restricting the naturalness, efficiency, and user experience of current interactions.

To allow AI agents to have the ability to predict the user's intention, previous works focused on predicting intentions based on Bayesian methods \cite{albrecht1998bayesian, baker2017rational, shergadwala2021can} and Deep Neural Networks (DNNs) \cite{casas2018intentnet, gebert2019end}. Bayesian-based methods can provide the uncertainty of the prediction but have the disadvantages of handling complex input data form (e.g. images) and gearing the probabilistic models for the domain they are trained in. DNN-based methods, on the other hand, are excellent at handling complex input data forms but cannot easily quantify the epistemic uncertainty in the prediction. 
A model that combines DNN-based and Bayesian-based methods together could have the advantages of both. 
This can benefit practical applications in two aspects. First, a collaborative AI agent needs to operate in the real world and it needs to deal with data with high (visual) complexity. Deploying DNNs can easily adapt to complex input. Second, the uncertainty quantification about the prediction of intention can help better provide decisions on future actions and reduce the risk of a potential wrong prediction. 

In this work, we propose a novel two-step procedure to infer human intentions from sequences of actions (\autoref{fig:teaser}).
Our approach combines DNNs to obtain the probabilities of the next action with MCMC-based Bayesian inference for inferring intentions.
Specifically, given action data from $N$ different intention, we train $N$ DNN models for next action prediction. At test time, one action sequence data is fed to all $N$ models to obtain the action probabilities, the action sequence represents the true intention. The output from $N$ DNNs represents the probabilities assuming $N$ intentions are applied. Next, we use Markov Chain Monte Carlo (MCMC) sampling to train a Bayesian model with all action probabilities from all DNNs to infer the intentions. Our two-step method decouples the next action prediction (DNNs) and actual intention prediction (Bayesian model). We do not have any requirements on the DNN input format and network architecture. They can be modified and optimised according to different tasks. The Bayesian model is independent of the DNNs, it only takes the action probabilities from DNNs and predicts the intentions with uncertainties. 
We demonstrate the effectiveness of our method through experiments on two datasets: Watch-And-Help (WAH)~\cite{puig2020watch} and keyboard and mouse interaction dataset~\cite{zhang22_caugaui}. 
Our results show that our method can correctly predict the intentions of users. We further evaluate the performance of our method with 10\% to 100\% of observed actions in one action sequence on both datasets. The results show that using 20\% of actions in a sequence, is often sufficient for the true intention to have clearly higher posterior probability than all other intentions, although with substantial uncertainty. This demonstrates that our method can infer the users' intention already in an early stage where only a few actions have been observed.

The main contribution of this work is the two-step method to infer human intention. We use action probabilities of next action prediction from DNNs with a Bayesian model to infer intentions. To our best knowledge, we are the first ones to propose the joint use of DNNs and Bayesian models to decouple the next action prediction and intention prediction. 
Our method has three advantages. First, the DNNs and the Bayesian inference are decoupled. The inference of intention does not depend on the DNN architecture. One can optimise the DNN architecture for classifying the next action separately. Second, training the Bayesian model requires less time and Bayesian inference provides a fast prediction on intention. Third, our method can predict the intention correctly and efficiently when using a few observed actions in the series of actions.

\section{Related Work}
\label{sec:related_work}

Human-AI collaboration has attracted increasing interest recently. Several works focused on developing computational agents for collaboration in virtual environments \cite{kolve2017ai2, szot2021habitat, puig2023nopa}. The virtual environments possess near-realistic scenes and objects and support different types of actions. 
The importance of intention prediction in collaboration has been shown in virtual environments \cite{puig2020watch} and real-life scenarios \cite{wang2013probabilistic}. Correctly predicted human intentions lead to more effective collaborations in robotic shared autonomy \cite{jain2019probabilistic}, Human-Robot handover \cite{wang2021predicting} and cooperative assembly \cite{liu2021unified}.
Several prior works have focused on action anticipation based on videos, i.e., the task of predicting future actions based on observed behaviour in the past \cite{gammulle2019predicting,furnari2020rolling, qi2021self}.
Different types of models have been used
, e.g. two-stream CNN~\cite{gammulle2019predicting}, LSTM \cite{furnari2020rolling}, video transformer~\cite{girdhar2021anticipative}, or graph neural networks~\cite{wu2021anticipating}. In \cite{camporese2021knowledge}, the authors further used label smoothing technique to improve the work \cite{furnari2020rolling}. Other works have also used goals in anticipating future actions \cite{chang2020procedure, roy2022action}. 


In \cite{huang2015using}, the intention was defined as the intended ingredients for a sandwich. SVMs were used to predict the intention from human gaze data. SVM was also used in intention prediction during human interactions~\cite{bednarik2012you}. Other approaches include MDP~\cite{koppula2016anticipatory}, probabilistic graphical model~\cite{song2013predicting}, k-nearest neighbour (kNN)~\cite{qu2019user}.
In \cite{zhang22_caugaui}, the authors investigated the task of predicting user intents from mouse and keyboard input as well as gaze behaviour.
In another line of work, gaze behaviour was also identified as a rich source of information for predicting users' search intents \cite{sattar15_cvpr,sattar17_iccvw,sattar20_neurocomp} and even visually reconstructing it \cite{strohm21_iccv}.
Perhaps the works in~\cite{singh2018combining, le2021goal} are the most similar to ours. In~\cite{singh2018combining}, a Bayesian model to infer intentions from ontic actions and gaze action. The ontic actions are the actions that change the state of the world and the gaze actions are the regions where an agent is looking at with regard to the ontic action. The work in~\cite{le2021goal} further introduced a deceptive component into the Bayesian model for the scenario where the human might perform ambiguous actions on purpose. Although Bayesian models were used for intention prediction, the actions were not predicted by neural networks. Rather they were pre-processed and then used in the Bayesian models.
\section{Method}
\label{sec:method}

For each of the $N$ possible intentions, we train a separate DNN on data where the ground-truth intention is known. The task of the DNNs is to predict the next action from all previous actions. 
Since our method works with arbitrary DNN architectures that perform this prediction task, we are not focusing on the specific architecture of the DNN here.
It is important, however, that each DNN has a final Softmax (or equivalent) layer to obtain the predicted next-action probabilities. All action probabilities are then used to train a Bayesian model to predict the intention from the set of predicted next-action probabilities (see below for details). At test time, the data of each intention is forwarded to all DNNs to obtain $N$ next-action probabilities representing $N$ intentions. 
We refer the action sequence forwarded to the DNNs with the known intention label as the true intention. The $N$ DNNs are trained with $N$ intentions and we interpret each DNN as an assumed intention, i.e. given one action sequence, the DNNs do not know the true intention, the $i$th DNN assumes it is from the $i$th intention. 
The Bayesian model then uses all action probabilities jointly to infer the posterior distribution over the $N$ assumed intention. Specifically, the Bayesian only uses the action probabilities from one action sequence to infer user intention. 
All DNNs can be trained separately as they do not need to share weights for our procedure to work.

Formally, an intention $\mathcal{I}$ consists of a series of actions,
\begin{equation}
    \mathcal{I}_{ij}=[a_0, ... , a_L], 0<i<N, 0<j<M, 0<k<L,
\end{equation}
where $a$ is action, $M$ is the number of action series belonging to $i^{th}$ intention and $L$ is the total number of actions in $\mathcal{I}_{ij}$. $I_{ij}$ represents an action series instance in the $i$th intention. In the training of $i$th DNN, we use all instances in $I_i$ as training data.
The input of for the DNN are the $\mathcal{I}_{ij}$, whereas the ground-truth next action $y_i=a_{k+1}$ constitutes the target variable. The loss function is then
\begin{equation}
    \mathcal{L}=f(y, \hat{y}),
\end{equation}
where $\hat{y}$ is the DNN prediction and $f(\cdot)$ is a cross entropy loss. After all networks are trained, each intention data is passed to all $N$ networks and obtains $N$ Softmax outputs. The $i^{th}$ Softmax outputs produced by the $i^{th}$ DNN represent the action probability assuming $i^{th}$ intention is applied.
To infer the intention of humans based on the series of actions and their DNN predictions serving as a surrogate likelihood, we set up the following Bayesian model:

\begin{align*}
\begin{split}
    a_k \sim \text{categorical}(\theta_k), \forall k=1,...,L, 
\end{split} \\
\begin{split}
    \theta_{km} = \sum_{i=1}^{N}P(a_{km}|\mathcal{I}_i)P(\mathcal{I}=\mathcal{I}_i), \forall m=1,...,M, 
\end{split} \\
\begin{split}
    P(\mathcal{I}) \sim \text{Dirichlet}(\alpha),
\end{split}
\end{align*}
where $P(\mathcal{I}=\mathcal{I}_i)$ is the $i^{th}$ element of the intention probability $P(\mathcal{I})$ to be inferred by the model, and $\alpha \in \mathbb{R}_+^{N}$ is the concentration vector of the Dirichlet prior on $P(\mathcal{I})$, which we set to $\alpha = 1$ to obtain an uninformative prior. The action probabilities $P(a_{km}|\mathcal{I}_i)$ of the $m$th possible action to occur at the $k$th position in the sequence are obtained from the output of DNNs. To predict the intention $P(\mathcal{I})$, we use the probabilistic programming language Stan \citep{carpenter2017stan}, which employs a state-of-the-art Markov-chain Monte-Carlo (MCMC) sampler to $P({\mathcal{I}})$.

\section{Experiment}
\label{sec:experiment}
\subsection{Datasets}

\subsubsection{Watch-And-Help Dataset}
WAH is a dataset for social intelligence and human-AI collaboration \cite{puig2020watch}. In the dataset, an AI agent Bob helps another human-like agent Alice perform household activities. The world is a 3D virtual environment. There are two stages of collaboration, i.e., the \textit{Watch} stage and the \textit{Help} stage. In the \textit{Watch} stage, Bob observes Alice demonstrating an activity and Bob helps Alice with the same activity in the \textit{Help} stage. 
In this work we only consider the \textit{Watch} state given that we are interested in inferring the intentions of Alice.
We understand the activities are defined by a set of sub-goals represented by predicates.
Both agents can perform different actions to accomplish their goals.
An activity is accomplished once the states of all sub-goal predicates are reached. In total, there are five types of activities with each activity having two to eight sub-goals.
The dataset has one training and two test sets.
To evaluate our method we only need information on the activity and actions and thus leave the sub-goals aside.
Furthermore, to keep the activity category consistent, we focus only on those types of activities that are present in the training set and test set 1.
We treat the activity as the intention and predict the intentions from the sequence of actions. Since we use the DNN to predict the next action in an action sequence, we modify the original action sequences for the use of next action prediction. For an action sequence $[a_0, ..., a_L]$, when a new action is observed, we create a new action sequence. The dataset does not have action sequences from different users, to evaluate from a user perspective, we create 92 artificial users and randomly assign action sequences to the users.

\subsubsection{Keyboard and Mouse Interaction Dataset}

To complement the household activities performed in the virtual environment in the WAH dataset, we also evaluated our method on keyboard and mouse interaction dataset introduced in \cite{zhang22_caugaui}.
16 participants were asked to format text according to several formatting rules (the interaction intentions). 
The evaluation task on this dataset was to predict these interaction intentions from mouse and keyboard input.
The text consisted of titles, subtitles and paragraphs and a rule contained instructions on how to format it using the mouse and keyboard (e.g. "make the title bold"). 
Participants could perform seven different actions for formatting the text.
The dataset contains data from two types of formatting tasks: 
First, participants were asked to perform formatting according to seven predefined formatting rules. Each rule was repeated five times.
Second, each participant was asked to create a custom rule themselves and to format the text according to this rule.
We only used data from the first part of the dataset for our experiment since there is only one intention for each participant in the second part.
We used the data from participants one to 11 for training and the data from participants 12 to 16 for testing.

\subsection{Experimental Settings}
We performed two experiments on the WAH dataset and the keyboard and mouse interaction dataset. 
We first evaluated our method on inferring users' intentions using the full action sequences. 
In the second experiment, we used 10\% to 90\% of the actions in an action sequence with a 10\% step to infer the intentions.
Since the WAH dataset is created in a virtual environment and the action sequences do not belong to any user, we created virtual users by randomly grouping the data in test set 1 and test set 2.
As a result, test set 1 had 92 artificial users, each user had one action sequence in \textit{put fridge}, two action sequences in \textit{put dishwasher}, and three action sequences in \textit{read book}.
Test set 2 had nine users, each user had one action sequence in \textit{put fridge}, five action sequences in \textit{put dishwasher}, and five action sequences in \textit{read book}. We show the results on test set 1 in \autoref{sec:result}, but we conducted experiments on both test set 1 and test set 2. We observed similar outcomes, we only show results on test set 1 due to the limit of space.
The architecture of our DNN model is based on the one in \cite{puig2023nopa}.

To train the DNNs on the WAH dataset, we used 2,000 epochs, a batch size of 32 and a learning rate of $3e^{-4}$. For the keyboard and mouse interaction dataset, we trained for 100 epochs, the batch size was eight, and the learning rate was $1e^{-4}$.
To train the Bayesian model we used the same training strategy for both datasets. 
For each action sequence, we performed Bayesian inference via four MCMC chains, each with 2,000 iterations of which the first 1,000 were discarded as warmup.
All Bayesian models converged well according to standard convergence criteria \cite{vehtari_rhat_2021}.

\section{Results}
\label{sec:result}

\subsection{User Intention Prediction}
\begin{figure*}[t]
    \centering
    \includegraphics[width=0.8\linewidth]{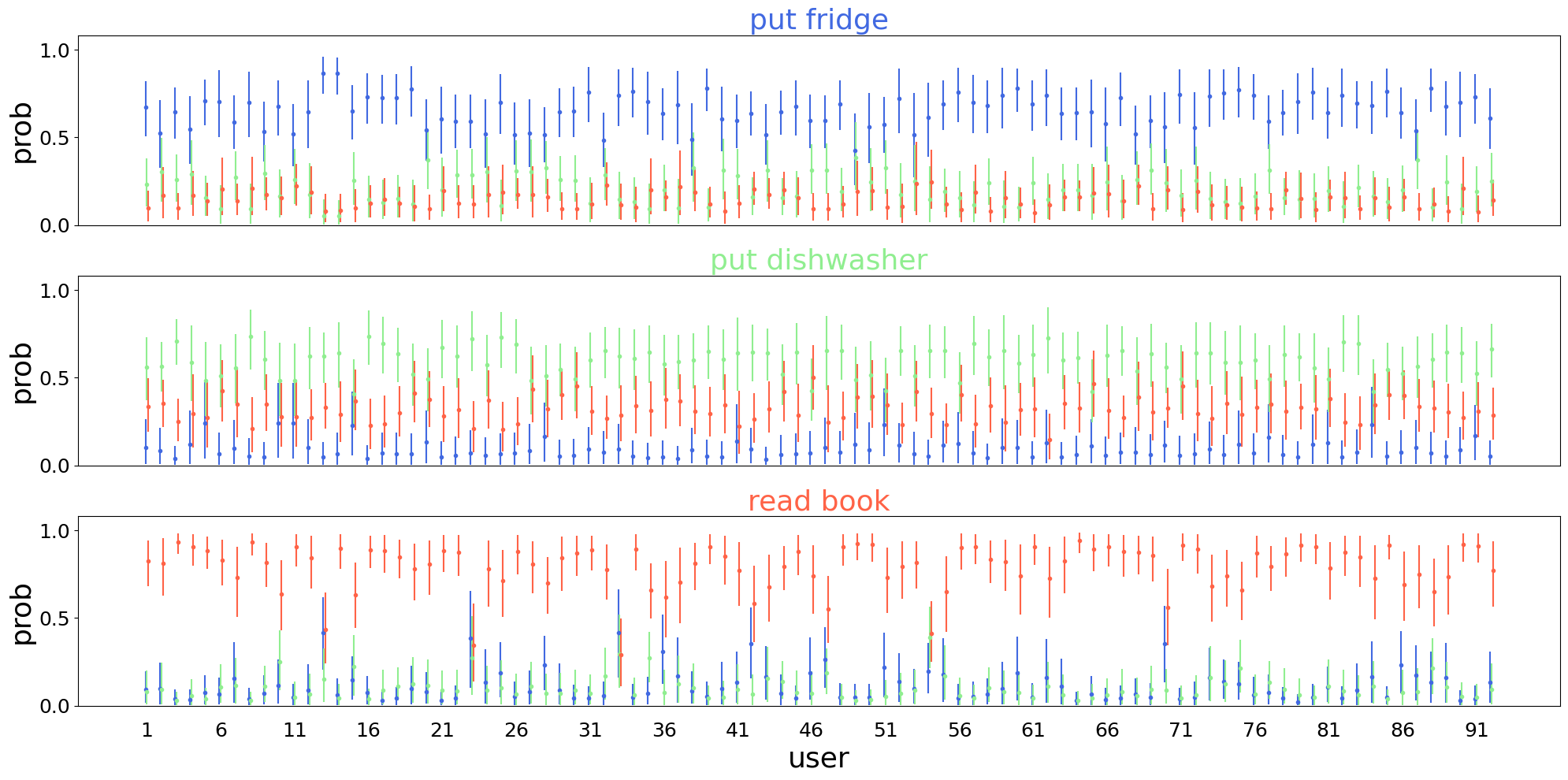}
    \caption{Result of intention prediction of users on test set 1 in WAH dataset.}
    \label{fig:intention_user_WAH_test_1}
\end{figure*}

\begin{figure*}[h]
    \centering
    \includegraphics[width=\linewidth]{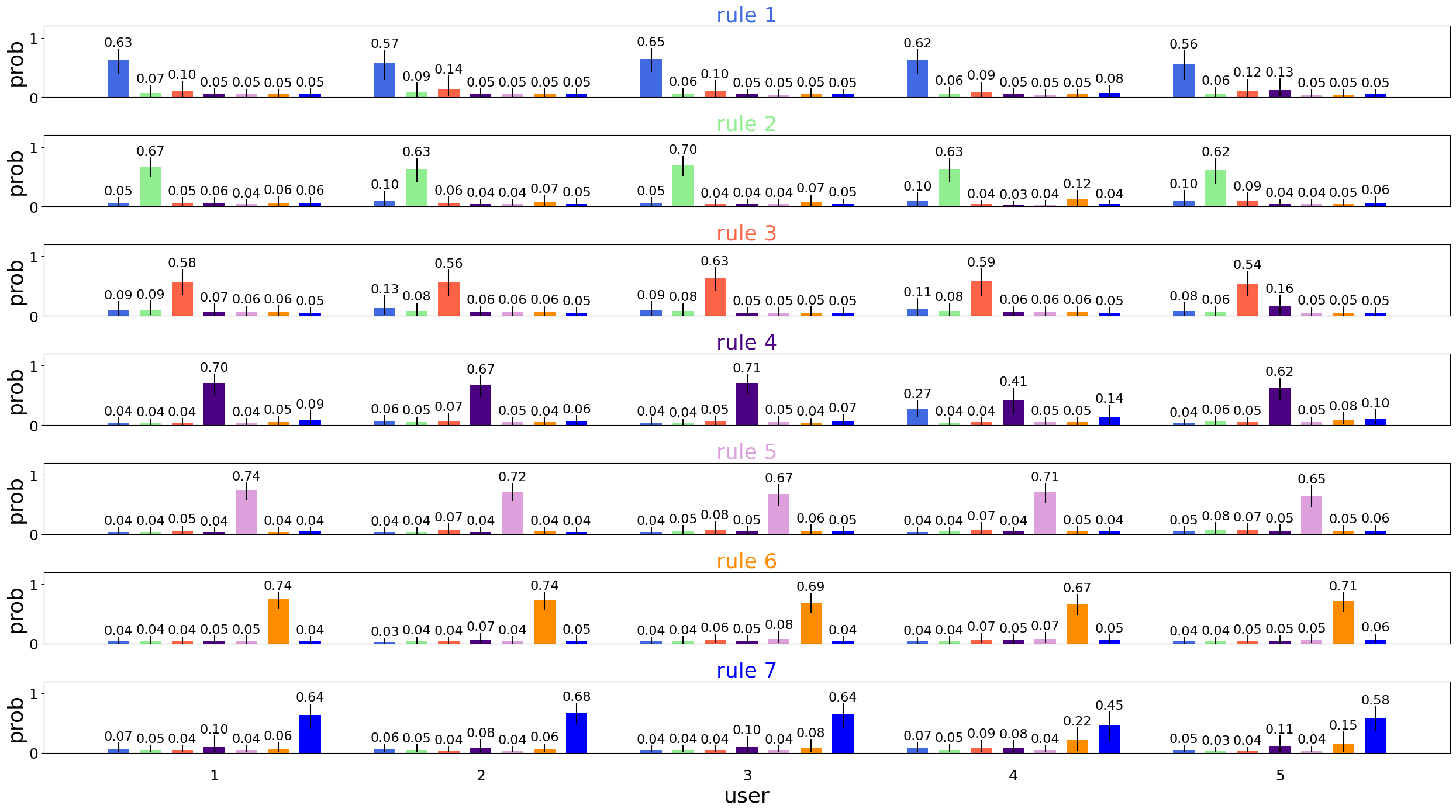}
    \caption{Result of intention prediction of users in keyboard and mouse interaction dataset.}
    \label{fig:intention_user_keyboard_mouse}
\end{figure*}

\autoref{fig:intention_user_WAH_test_1} shows the result of user intention prediction on test set 1 of the WAH dataset. We report the posterior mean and 90\% credible intervals (CIs) of the probabilities of all assumed intentions.
The top, middle and bottom plot shows the results when the true intention is \textit{put fridge}, \textit{put dishwasher} and \textit{read book}. 
For the true intention \textit{put fridge}, for most users our method can predict the correct intention, meaning that the assumed intention with the highest posterior mean probability is the same as the true intention. In a few cases, the posterior mean probability of put fridge is close to put dishwasher or read book.
We can see that the posterior mean of put fridge and put dishwasher for user 49 is 0.43 and 0.39. 
When the true intention is \textit{read book}, our method can also predict the correct intentions of most users with a few exceptions,
i.e. user 13, 23, and 33. For user 13, the posterior mean of read book 0.43, only 0.02 higher than put fridge. For user 23 and 33, the posterior mean of put fridge is slightly higher than read book. 
For the true intention \textit{put dishwasher}, although the posterior mean of put dishwasher are the highest in most users predictions, the difference between put dishwasher and the other two assumed intentions are smaller compared to the cases in true intention \textit{put fridge} and \textit{read book}. 
For user 6, 19, 20, 44, 49, 51, 53, the difference between the posterior mean of put dishwasher and the posterior mean of read book are around 0.1. For user 15, 27, 30, 46, 50, 56, 65, 71, and 84, the differences are below 0.07.
Overall, our model can predict the correct true intention \textit{put fridge}, however the prediction for user 49 is rather uncertain. For true intention \textit{read book}, predictions for most users are correct but more predictions are more uncertain. The model can predict users' true intention \textit{put fridge} and \textit{read book} better than \textit{put dishwasher}.

\autoref{fig:intention_user_keyboard_mouse} shows the posterior mean and 90\% CIs of the predicted intentions in the keyboard and mouse interaction dataset. The prediction on all user data on all rules are correct in terms of the highest posterior mean of the assumed intention being the true intention. 
For user 1, 2,3 and 5, the differences of the posterior mean between the correctly predicted intention and the rest intentions in all true intentions are quite large. 
For user 4, the posterior of the correct intention for true intention \textit{rule 4} and \textit{rule 7} are more uncertain than the other true intentions. For true intention \textit{rule 4}, the posterior mean of rule 4 is 0.41 while the posterior mean of rule 1 is 0.27. For true intention \textit{rule 7}, the posterior mean probabilities of rule 7 and rule 6 are 0.45 and 0.22 respectively.

\begin{figure*}[h]
    \centering
    \includegraphics[width=0.8\linewidth]{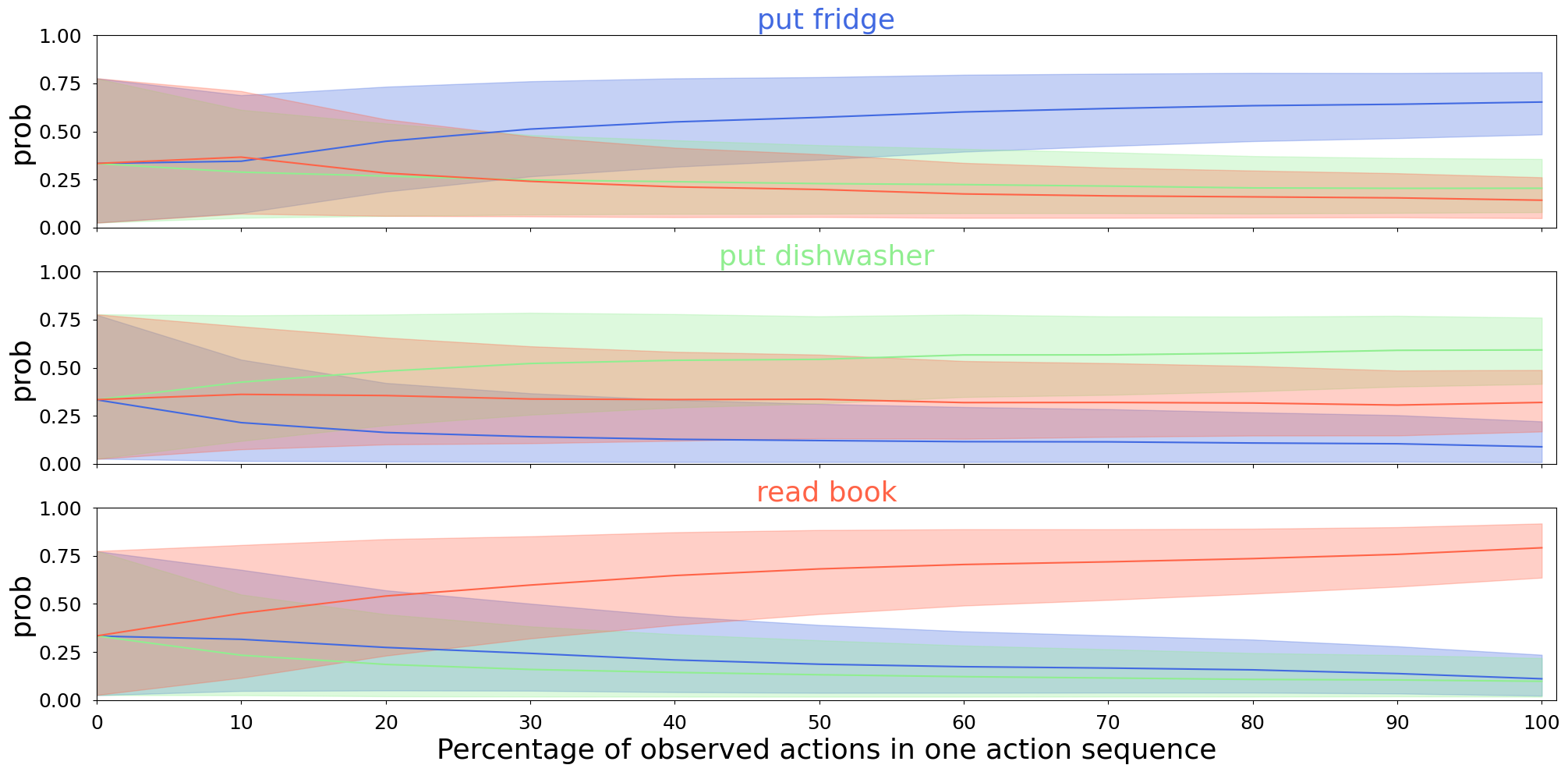}
    \caption{Posterior mean probabilities and CI bounds when different percentages of observed actions in an action sequence are used for inference. The results on test set 1 in WAH dataset are shown.}
    \label{fig:act_seq_WAH_test_set_1}
\end{figure*}

\begin{figure*}[h]
    \centering
    \includegraphics[width=0.8\linewidth]{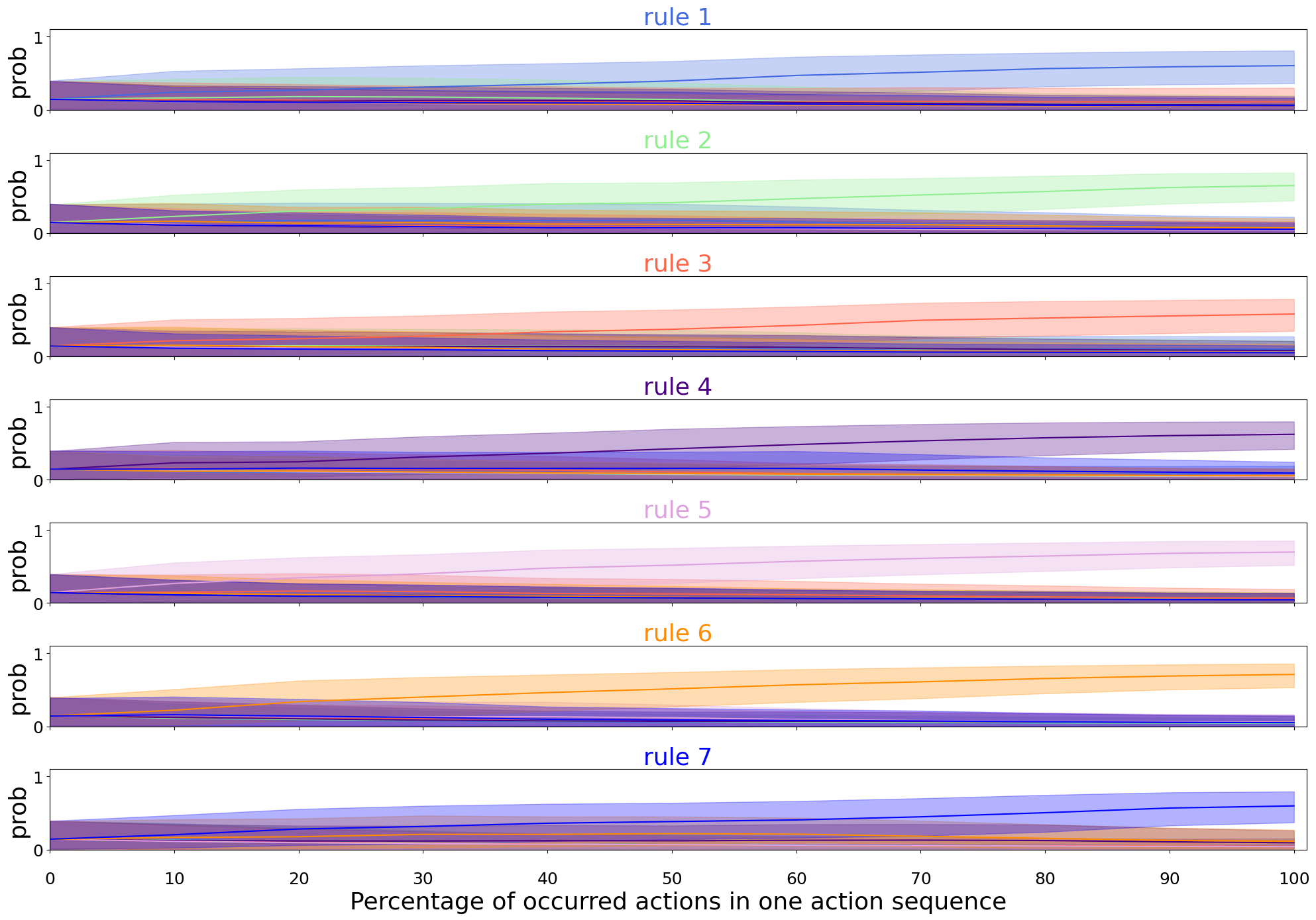}
    \caption{Posterior mean probabilities and CI bounds when different percentages of observed actions in an action sequence are used for inference. The results in keyboard and mouse interaction dataset are shown.}
    \label{fig:act_seq_keyboard}
\end{figure*}

\subsection{Different Lengths of Observed Actions in Action Sequences}

\autoref{fig:act_seq_WAH_test_set_1} shows the results when different percentages of observed actions in one action sequence are used for inferring intention on test set 1 in WAH dataset. The percentage of the action in a sequence is varied from 10\% to 100\% in steps of 10\%. For instance, 50\%, we extract the first 50\% of the actions in the sequence and use it to infer the intention. We plot the average posterior mean probabilities and 90\% CIs, i.e. at each percentage of observed actions in one sequence, the values of posterior mean and CI bounds are the average values from all users. 
For all true intentions, the posterior mean probabilities of the correct intentions are already relatively higher than the other assumed intentions when 20\% of action in a sequence has been observed.
The posterior mean probabilities increase with the increase of observed actions in action sequences. The CI bounds decrease as the percentage of observed actions increases. This shows that the more actions have been observed in one action sequence, the more certain the Bayesian model is about its intention predictions. 

When the true intention is \textit{put dishwasher}, the predictions of Bayesian models are more uncertain than the other two true intentions. This can be observed when predicting user intention using full action sequence (\autoref{fig:intention_user_WAH_test_1}) and partially observed action sequence (\autoref{fig:act_seq_WAH_test_set_1}). 
We interpret that it is due to the noisier distribution of the actions in action sequences and the predictions of the DNNs. 
That is, the actions in the action sequences are not representative enough. By representative actions we mean the actions from which the intention can be easily interpreted. For instance, \textit{open fridge} is a representative action for the intention \textit{put fridge}.

\autoref{fig:act_seq_keyboard} shows the result on the keyboard and mouse interaction dataset. We show the average posterior mean probabilities and the 90\% CIs of all seven intentions when different percentages of actions in one action sequence are used for inference. 
For all true intentions, the posterior mean probabilities of the correct intentions increase with more actions having been observed in the action sequences.
At 10\%, the assumed intention with the highest posterior mean probability is the same as the true intention for all rules but the differences are small. The predictions are still quite uncertain. At 20\%, the differences in true intention \textit{rule 2}, \textit{rule 5}, and \textit{rule 6} become larger, but the CI bounds remain at wide ranges.
For true intentions \textit{rule 5} and \textit{rule 6}, the probabilities of correct intentions are close to 0.5, however, the CIs have not decreased a lot. At 50\%, the CIs of the correct intention already no longer overlap with the CIs of the other intentions. 
\section{Discussion}
\label{sec:discussion}

In the evaluation of WAH dataset, we manually created artificial users by assigning action sequences to users and most intentions of users were inferred correctly. 
In the keyboard and mouse interaction dataset, all predictions of all users for all true intentions were correct.
We used one action sequence from one user to perform Bayesian inference. This shows the Bayesian model is efficient for inferring intention in terms of the number of observations of action sequence. It does not have to see multiple action sequences to infer the correct intention, only seeing one action sequence is adequate. 

We were also interested in how the Bayesian model performs when fewer actions have been observed in the action sequences. 
An intuition is that the model is more confident about the inferred intention when more actions have been observed. This is confirmed by experiments in two aspects. First, the posterior mean probabilities increase when more actions are observed. Second, the ranges of CI bounds become smaller meaning the Bayesian model is more certain about its predictions. 
Additionally, the Bayesian model can predict the true intentions correctly even at an early stage in the action sequence.
Being able to predict human/agent intention in an early stage can benefit agent-agent and human-agent interaction.
For instance, in the WAH scenario, an agent can help the other agent finish a task by completing other actions in the same task. In the scenario of keyboard and mouse interaction, the computer/agent can optimise the user interface or give suggestions while the human is formatting the text. 
It is necessary that the agent has enough time to plan and deploy collaboration and interaction. To be able to predict intentions when only partial actions in an action sequence allows the agent to have sufficient time for planning.
It is worth noting that the uncertainties of the predicted intention in early stages are relatively high and this
should be taken into consideration when designing the interaction with a human. 

When the number of intentions scales up, we can train the DNN in a multi-task setting, i.e. each DNN representing an intention is jointly trained and they do not share weights. This is equivalent to training separate DNNs but saves the time of training. As for the Bayesian inference, more intentions would not affect the computational time significantly.

\section{Conclusion}
\label{sec:conclusion}

In this work we proposed a two-step procedure to infer human intentions from a series of actions based on DNNs and Bayesian inference.
First we trained DNNs to obtain the probabilities of predicted next action in a sequence.
Then we used MCMC-based Bayesian inference to infer the human intention from the predicted next-action probabilities.
We performed experiments on the WAH and keyboard and mouse interaction datasets to validate our approach.
The results show that we can accurately infer the intentions even when only one action sequence from one user is available at inference time.
This suggests that the implicit information contained in the next action probabilities generated by DNNs can be used to infer the intention using a Bayesian model.
In addition, we demonstrated that our approach still provides correct predictions even if only a few actions have been observed.

\section{Acknowledgement}
Lei Shi is funded by the Deutsche Forschungsgemeinschaft (DFG, German Research Foundation) under Germany's Excellence Strategy -- EXC 2075 -- 390740016. 
Andreas Bulling is funded by the European Research Council (ERC) under the European Union's Horizon 2020 research and innovation programme under grant agreement No 801708.

\bibliographystyle{ACM-Reference-Format}
\bibliography{ref}


\end{document}